\documentclass[preprint,superscriptaddress,amsmath,amssymb,aps,notitlepage,floatfix,longbibliography]{revtex4-2}

\usepackage{graphicx}
\usepackage{dcolumn}
\usepackage{bm}
\usepackage{color}
\usepackage{comment}
\graphicspath{{./Figures_JAP}}

\usepackage[utf8]{inputenc}
\usepackage[T1]{fontenc}
\usepackage{mathrsfs}
\usepackage{etoolbox}

\usepackage{upgreek}
\usepackage{gensymb}
\usepackage{float}
\usepackage{xcolor}
\usepackage{physics}
\usepackage{multirow}
\usepackage{array}
\usepackage{here}

\makeatletter
\def\@email#1#2{%
	\endgroup
	\patchcmd{\titleblock@produce}
	{\frontmatter@RRAPformat}
	{\frontmatter@RRAPformat{\produce@RRAP{*#1\href{mailto:#2}{#2}}}\frontmatter@RRAPformat}
	{}{}
}%
\makeatother
\begin{document}


\title[]{Model calculations of the strains associated with surface acoustic waves}

\author{Takuya Kawada}
\email[]{takuyakawada@g.ecc.u-tokyo.ac.jp}
\affiliation{Department of Physics, The University of Tokyo, Bunkyo, Tokyo 113-0033, Japan}
\affiliation{Department of Basic Science, The University of Tokyo, Meguro, Tokyo 152-8902, Japan}

\author{Masashi Kawaguchi}
\affiliation{Department of Physics, The University of Tokyo, Bunkyo, Tokyo 113-0033, Japan}

\author{Hiroki Matsumoto}
\affiliation{Department of Physics, The University of Tokyo, Bunkyo, Tokyo 113-0033, Japan}
\affiliation{Institute for Chemical Research, Kyoto University, Uji, Kyoto 611-0011, Japan}

\author{Masamitsu Hayashi}
\email[]{hayashi@phys.s.u-tokyo.ac.jp}
\affiliation{Department of Physics, The University of Tokyo, Bunkyo, Tokyo 113-0033, Japan}
\affiliation{Trans-Scale Quantum Science Institute (TSQS), The University of Tokyo, Bunkyo, Tokyo 113-0033, Japan}

\date{\today}

\begin{abstract}
	Magnon-phonon coupling has garnered increasing interest in condensed matter physics due to its fertile physics and potential applications in devices with novel functionalities.
	Surface acoustic waves (SAWs) are commonly employed as a source of coherent acoustic phonons.
	The strain associated with SAWs couples to magnetization of magnetic materials via magnetoelastic coupling and/or spin-rotation coupling.
	A typical SAW device is formed on a piezoelectric substrate with anisotropic crystal structure.
	Since the form of strain depends on the material parameters and structure of the SAW device, it is of vital importance to understand its character.
	In this paper, we present a comprehensive methodology to numerically calculate the SAW velocity, SAW excitation efficiency, lattice displacement and all strain components associated with SAW. LiNbO$_3$ is used as a prototypical material system.
	All quantities depend on the SAW propagation direction with respect to the crystalline axis and on the electrical boundary conditions.
	In contrast to non-piezoelectric isotropic media, we find that all shear strain components can be induced in LiNbO$_3$, with their amplitude and relative phase (with respect to the longitudinal strain) dependent on the propagation direction and the boundary conditions at the LiNbO$_3$ surface.
	These results offer a robust foundation for analyzing strain-driven magnon-phonon coupling mechanisms and contribute to designing strain-engineered functional magnonic and phononic devices.
\end{abstract}

\maketitle

\section{Introduction}

Spin waves~\cite{bloch1930,vanvleck1945rmp,herring1951pr} are collective spin precessions propagating in magnetic materials. Due to the small energy dissipation
while they travel, spin waves are considered as potential information carriers in next-generation memory and computation devices~\cite{kostylev2005apl,chumak2015np}. 
The quanta of spin waves, often referred to as magnons, are known to interact with other quasiparticles in solids. 
Such hybrid quasiparticles can acquire additional functionalities, attractive for developing novel devices~\cite{soykal2010prl,bai2015prl,chen2018prl,an2020prb,costa2023nl}.
Phonons, quanta of vibration modes of atoms, can couple to magnons via the magnetoelastic coupling~\cite{weiler2011prl,thevenard2014prb,gowtham2015jap} and/or the spin-rotation coupling~\cite{kurimune2020prl,tateno2021prb}. 
Phonons and magnons can be resonantly coupled, garnering interest in recent magnonic, phononic, and spintronic studies.
Studies reported magnon-phonon coupling strength of the order or tens of MHz up to SAW frequencies of few GHz~\cite{hatanaka2022prapp,matsumoto2024nl,hwang2024prl}.
Thanks to the coupling, a number of intriguing phenomena, e.g. acoustic spin pumping~\cite{weiler2012prl,xu2018prb} and non-reciprocial transport of phonons~\cite{sasaki2017prb,kuss2020prl,xu2020sciadv,piyush2020sciadv,tateno2020prapp,matsumoto2022apex,liyang2024sciadv}, were found.

Surface acoustic waves (SAWs), acoustic phonons localized at the surface of a solid, are often employed to study the magnon-phonon coupling. SAW can be excited by applying an rf signal to an interdigital transducer (IDT) on a piezoelectric substrate~\cite{white1965apl}. 
Understanding the SAW modes is of fundamental importance because the magnetoelastic effective field depends on the strains~\cite{weiler2011prl,dreher2012prb}.
The SAW mode for a piezoelectric medium shows large anisotropy due to its low crystal symmetry.
As a consequence, the SAW propagation characteristics (e.g. strain, stress) strongly depends on the direction to which it travels~\cite{datta1986book}. 
To develop functional magnon-phonon hybrid systems, it is essential to determine the strain components for SAWs that travel in different directions.

In this work, we present a comprehensive approach to calculating the strain components of SAWs propagating on the surface of piezoelectric substrates.
LiNbO$_3$ is selected as a model system due to its frequent application in magnon-phonon coupling studies.
Using material parameters of LiNbO$_3$, we numerically calculate the amplitude and phase of the SAW strains as a function of its propagation direction with respect to the crystal coordinate axis. 
We find that all shear strain components can be generated owing to the anisotropy of the substrate. 
In addition, we show that the electrical boundary conditions at the substrate surface can influence the magnitude of the strains. 

\section{Methods}
\subsection{\label{sec:modeldescription}Model description}
\begin{figure}[b]
	\begin{minipage}{1.0\hsize}
		\centering
		\includegraphics[scale=0.08]{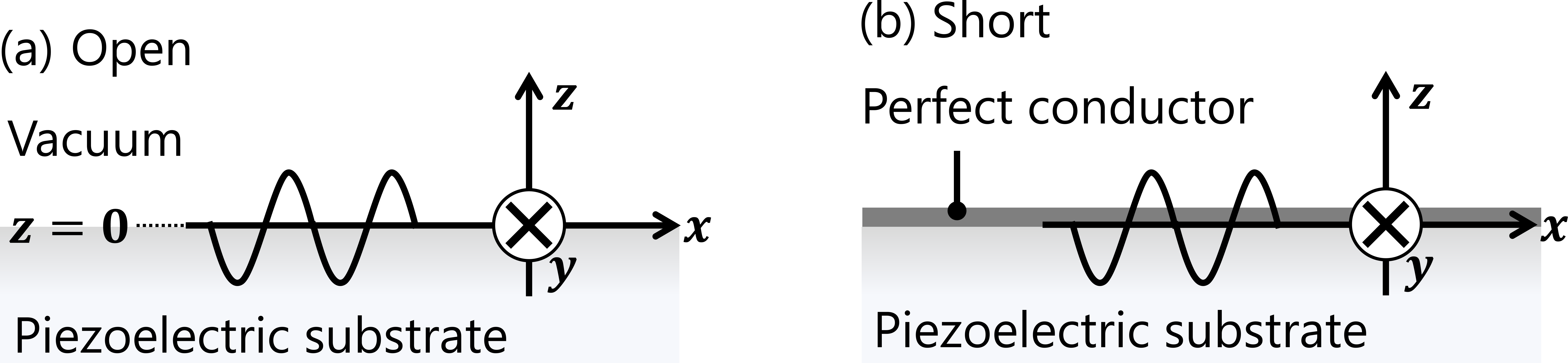}
	\end{minipage}
	\caption{(a,b) Schematic illustration of the lab coordinate system. The system consists of a piezoelectric substrate/vacuum (a) and piezoelectric substrate/perfect conductor/vacuum (b), respectively. The thickness of the perfect conductor is assumed to be infinitesimally thin. $x$- and $z$-axis are set along the SAW propagation direction and surface normal.  
		\label{fig:Fig1}
	}
\end{figure}

The coordinate axis of the system is described in Fig.~\ref{fig:Fig1}. The substrate normal is along the $z$-axis and the SAW propagates along the $x$-axis, i.e. along the surface of the substrate. We assume that all strain components of SAW are uniform along the $y$-axis. 
In the following, indices $i,j,k,l = 1,2,3$ indicate the spatial coordinates with the following convention: $1\leftrightarrow x$, $2\leftrightarrow y$, and $3\leftrightarrow z$. The Einstein convention, summation over repeated indices, is implied.

\subsection{\label{sec:eom}SAW equations of motion}
First, we describe the governing equations that determine the dynamics of SAW in a piezoelectric substrate. 
The governing equations consist of the equation of motion for elastic waves and the Poisson equation:
\begin{equation}
	\begin{aligned}
		\rho_\mathrm{p} \pdv[2]{u_i}{t} = \frac{\partial T_{ij}}{\partial x_j},
		\label{eq:eom:piezo:u}
	\end{aligned}
\end{equation}
\begin{equation}
	\begin{aligned}
		\div{\bm{D}}=0.
		\label{eq:eom:piezo:D}
	\end{aligned}
\end{equation}
where $\bm{u}$ is the lattice displacement vector, $T_{ij}$ is the stress, $\bm{D}$ is the displacement field and $\rho_\mathrm{p}$ is the mass density of the substrate.
In almost all cases~\cite{tiersten1963jasa,tseng1967jap,campbell1968ieee,ingebrigtsen1969jap}, the electrostatic approximation,
\begin{equation}
	\begin{aligned}
		\bm{E} = - \nabla \phi,
		\label{eq:electrostatic}
	\end{aligned}
\end{equation}
is invoked to solve the equations of motion, where $\bm{E}$ is the electric field and $\phi$ is the scalar potential.
The effect of magnetic field is thus neglected. 
In piezloelectric materials, the stress $T_{ij}$ is expressed with a linear combination of the symmetrized strain $S_{kl}\equiv(\partial_j u_i + \partial_i u_j)/2$ and the electric field:
\begin{equation}
	\begin{aligned}
		T_{ij} &= c_{ijkl}S_{kl}-e_{ijk} E_k.
		\label{eq:piezo_eq_fund:T}
	\end{aligned}
\end{equation}

The elastic constant $c_{ijkl}$ is defined under a constant electric field and $e_{ijk}$ is the piezoelectric constant under a constant strain. From the symmetric constraints, $c_{ijkl}=c_{jikl}=c_{ijlk}=c_{klij}$ and $e_{ijk}=e_{jik}=e_{jki}$ are satisfied in many systems that include piezoelectric materials.
Similarly, the displacement field $D_i$ is given by a linear combination of the symmetrized strain and the electric field:
\begin{equation}
	\begin{aligned}
		D_i &= e_{ijk} S_{jk} + \varepsilon_{ij} E_j.
		\label{eq:piezo_eq_fund:D}
	\end{aligned}
\end{equation}
The dielectric constant $\varepsilon_{ij}$ is defined under constant strain.

Substituting Eqs.~(\ref{eq:piezo_eq_fund:T}) and (\ref{eq:piezo_eq_fund:D}) into Eqs.~(\ref{eq:eom:piezo:u}) and (\ref{eq:eom:piezo:D}), respectively, we obtain equations that relate the lattice displacement with the scalar potential: 
\begin{equation}
	\begin{aligned}
		\rho_p \pdv[2]{u_i}{t}=c_{ijkl}\pdv[2]{u_k}{x_j}{x_l}+e_{ijk}\pdv[2]{\phi}{x_j}{x_k}.
		\label{eq:eom}
	\end{aligned}
\end{equation}
\begin{equation}
	\begin{aligned}
		e_{ijk}\pdv[2]{u_j}{x_i}{x_k}-\varepsilon_{ij}\pdv[2]{\phi}{x_i}{x_j}=0.
		\label{eq:poisson}
	\end{aligned}
\end{equation}

In Eqs.~(\ref{eq:eom}) and (\ref{eq:poisson}), we have four unknowns: $u_i$ ($i = x, y, z$) and $\phi$.
The general form of the plane-wave solutions to Eqs.~(\ref{eq:eom}) and (\ref{eq:poisson}) are written as
\begin{equation}
	\begin{aligned}
		\bm{U}=\tilde{\bm{U}} e^{i q \kappa z} e^{i q\qty(x - v t)},
		\label{eq:plane}
	\end{aligned}
\end{equation}
where
\begin{equation}
	\begin{aligned}
		\bm{U}\equiv \mqty[u_x & u_y & u_z & \phi]^T, \ \ \tilde{\bm{U}} \equiv \mqty[\tilde{u_x} & \tilde{u_y} & \tilde{u_z} & \tilde{\phi}]^T.
		\label{eq:plane_abb}
	\end{aligned}
\end{equation}
The superscript $[...]^T$ represents transpose operation.
$q$ and $v$ are the in-plane wavenumber and the SAW velocity, which sets the angular frequency as $\omega = qv$.
$\kappa$ represents the wavenumber along $z$ normalized by $q$.
Note that we assume SAW is uniform along $y$ and thus Eq.~(\ref{eq:plane}) is independent of $y$.
Substituting Eq.~(\ref{eq:plane}) into Eqs.~(\ref{eq:eom}) and (\ref{eq:poisson}) gives the following matrix form eigenequation:
\begin{widetext}
\begin{equation}
	\begin{aligned}
	\mathcal{M} \tilde{\bm{U}}&=\bm{0},\\
	\mathcal{M} &=\mqty[ c_{1i1j}k_i k_j - \rho_\mathrm{p} v^2 &c_{1i2j}k_i k_j&c_{1i3j}k_i k_j&e_{1ij} k_i k_j \\
		c_{1i2j}k_i k_j &c_{2i2j}k_i k_j - \rho_\mathrm{p} v^2 &c_{2i3j}k_i k_j&e_{2ij} k_i k_j \\ 
		c_{1i3j}k_i k_j &c_{2i3j}k_i k_j &c_{3i3j}k_i k_j - \rho_\mathrm{p} v^2&e_{3ij} k_i k_j \\
		e_{1ij} k_i k_j & e_{2ij} k_i k_j & e_{3ij} k_i k_j &-\varepsilon_{ij} k_i k_j ].
		\label{eq:poi_sol2}
	\end{aligned}
\end{equation}
\end{widetext}
Here we used the relations $e_{ijk}=e_{jik}$, $c_{ijkl}=c_{klij}$ and set $\bm{k}=(1,0,\kappa)$ to derive Eq.~(\ref{eq:poi_sol2}).
Note that $q\bm{k}=(q,0,q\kappa)$ is the complex wave vector of the SAW (see Eq.~(\ref{eq:plane})).
A non-trivial eigenvector $\tilde{\bm{U}}\neq \bm{0}$ exists only when the determinant of $\mathcal{M}$ vanishes:
\begin{equation}
	\begin{aligned}
		\mathrm{det} [\mathcal{M}]=0.
		\label{eq:detm_zero}
	\end{aligned}
\end{equation}
Equation~(\ref{eq:detm_zero}) is an eighth order equation about $\kappa$ for a given value of $v$.
The eight solutions consist of four pairs with each pair being complex conjugate to each other.
We require the solutions to decay when $z \rightarrow -\infty$. 
To fulfill this condition, the four solutions of $\kappa$ whose imaginary part is negative are selected (note that $z < 0$ for the region inside the substrate).
Such solutions are defined as $\kappa_\alpha$ ($\alpha=1$--$4$) and the corresponding eigenvectors $\tilde{\bm{U}}$ that satisfy Eq.~(\ref{eq:poi_sol2}) are denoted as $\tilde{\bm{U}}_\alpha$.
The general solution is represented by the linear combination of the four solutions:
\begin{equation}
	\begin{gathered}
		\bm{U}=\qty(\sum_{\alpha=1}^{4} G_\alpha \tilde{\bm{U}}_\alpha e^{i q \kappa_{\alpha} z})e^{i q (x - vt)},
		\label{eq:gen_sol}
	\end{gathered}
\end{equation}
where $G_\alpha$ is the coefficient determined to satisfy the boundary conditions.

\subsection{Boundary conditions}
To determine $G_\alpha$, the boundary conditions need to be specified.
Since the boundary conditions at the substrate surface ($z=0$) influence $\bm{u}$ and $\phi$, here we examine two cases: when the substrate surface is (a) in contact with vacuum or (b) covered by an infinitesimally thin perfect conductor (referred to as a film hereafter), which are referred to as open and shorted boundaries, respectively.
At $z=0$, we have mechanical and electrical boundary conditions.
For the mechanical boundary condition, we assume a stress-free boundary at $z=0$ regardless of what covers the substrate surface, that is,
\begin{equation}
	\begin{aligned}
		T_{i3} \big|_{z=0} = \left[ c_{i3kl}\pdv{u_k}{x_l}+e_{i3k}\pdv{\phi}{x_k} \right]_{z=0} = 0.
		\label{eq:strain_bound}
	\end{aligned}
\end{equation}
The boundary condition~(\ref{eq:strain_bound}) states that the stress normal to the substrate surface vanishes at the surface ($z=0$).
Strictly speaking, this condition does not exactly hold when a film is placed on the substrate surface.
However, it is a rather good approximation when the film mass on the SAW propagation line is sufficiently small. 

The electrical boundary condition at the substrate surface ($z=0$) is given by the following form~\cite{ingebrigtsen1969jap}:
\begin{equation}
	\begin{aligned}
		\left. \frac{\phi}{D_z} \right|_{z=0} = -i \frac{v^2}{\omega}Z_\mathrm{p}.
		\label{eq:surf_imp}
	\end{aligned}
\end{equation}
$Z_\mathrm{p}$, referred to as the surface impedance, takes different forms depending on the boundary condition.
The value of $Z_\mathrm{p}$ for cases (a) and (b) is given as follows:
(a) substrate/vacuum (open boundary condition)
\begin{equation}
	\begin{aligned}
		Z_\mathrm{p}=\frac{i}{\varepsilon_0 v},
		\label{eq:Zp:open}
	\end{aligned}
\end{equation}
(b) substrate/infinitesimally thin perfect conductor (shorted boundary condition)
\begin{equation}
	\begin{aligned}
		Z_\mathrm{p}=0,
		\label{eq:Zp:open}
	\end{aligned}
\end{equation}
where $\varepsilon_0$ is the vacuum permittivity.

Given the boundary conditions (Eqs.~(\ref{eq:strain_bound}) and (\ref{eq:surf_imp})), we look for the solutions of the equations of motion. Substituting the plane-wave solution~(\ref{eq:gen_sol}) into Eqs.~(\ref{eq:piezo_eq_fund:T}) and (\ref{eq:piezo_eq_fund:D}) return the form of $T_{ij}$ and $D_i$. 
$T_{ij}$ and $D_i$ are then substituted into the boundary conditions~(\ref{eq:strain_bound}) and (\ref{eq:surf_imp}), which give
\begin{equation}
	\begin{aligned}
		\mathcal{N} \bm{G} = \bm{0},
		\label{eq:plane_bound_3}
	\end{aligned}
\end{equation}
where
\begin{equation}
	\begin{aligned}
		\mathcal{N} \equiv \mqty[\bm{N}_1 & \bm{N}_2&\bm{N}_3& \bm{N}_4], \ \  \bm{G} \equiv \mqty[G_1 & G_2 & G_3 & G_4]^T.
		\label{eq:plane_bound_3_Kvec}
	\end{aligned}
\end{equation}
$\bm{N}_{\alpha = 1-4}$ is a four-component vector that reads
\footnotesize
\begin{equation}
	\begin{aligned}
		\bm{N}_\alpha=\mqty[
		c_{131i} k_{\alpha ,i}  &c_{132i} k_{\alpha ,i} &c_{133i} k_{\alpha ,i} & e_{13i}k_{\alpha ,i}\\
		c_{231i} k_{\alpha ,i}  &c_{232i} k_{\alpha ,i} &c_{233i} k_{\alpha ,i} & e_{23i}k_{\alpha ,i}\\
		c_{331i} k_{\alpha ,i}  &c_{332i} k_{\alpha ,i} &c_{333i} k_{\alpha ,i} & e_{33i}k_{\alpha ,i}\\
		vZ_\mathrm{p} \cdot e_{13i}k_{\alpha ,i} & vZ_\mathrm{p} \cdot e_{23i}k_{\alpha ,i} & vZ_\mathrm{p} \cdot e_{33i}k_{\alpha ,i} & -1 - vZ_\mathrm{p} \cdot \varepsilon_{3i} k_{\alpha ,i}
		]\tilde{\bm{U}}_\alpha.
		\label{eq:plane_bound_2}
	\end{aligned}
\end{equation}
\normalsize
Here we defined $\bm{k_{\alpha}} = (1, 0, \kappa_{\alpha})$.
A non-trivial plane-wave solution ($\bm{G}\neq \bm{0}$) exists when 
\begin{equation}
	\begin{aligned}
		\mathrm{det} [\mathcal{N}] = 0.
		\label{eq:detn_zero}
	\end{aligned}
\end{equation}
If Eq.~(\ref{eq:detn_zero}) is satisfied, then $\bm{G}$ is obtained by solving the eigenequation~(\ref{eq:plane_bound_3}) and the general solution is constructed using Eq.~(\ref{eq:gen_sol}).


\subsection{Numerical calculation procedure}
Until now, we have assumed that the SAW velocity $v$ is \textit{a priori} given. However, $v$ is also an unknown, particularly for anisotropic piezoelectric materials, where $v$ depends on the direction to which the SAW travels.
To determine $v$, $\kappa_\alpha$, $\tilde{\bm{U}}_\alpha$ and $\bm{G}$, we take the following approach.
First, an arbitrary value of $v$ is substituted into the matrix $\mathcal{M}$ in Eq.~(\ref{eq:poi_sol2}).
We solve Eq.~(\ref{eq:detm_zero}) to obtain $\kappa_\alpha$ and the corresponding eigenvector $\tilde{\bm{U}}_\alpha$.
Next, $v$, $\kappa_\alpha$ and $\tilde{\bm{U}}_\alpha$ are substituted into Eq.~(\ref{eq:plane_bound_2}) to calculate $\mathrm{det} [\mathcal{N}]$.
If $\abs{\mathrm{det} [\mathcal{N}]}$ is sufficiently close to zero, $v$, $\bm{\kappa}_\alpha$ and $\tilde{\bm{U}}_\alpha$ are the correct solution and $\bm{G}$ is obtained from Eq.~(\ref{eq:plane_bound_3}).
When $\abs{\mathrm{det} [\mathcal{N}]}$ is not sufficiently close to zero, $v$ is slightly changed and the sequence is iterated until $\abs{\mathrm{det} [\mathcal{N}]}$ converges to zero.
From this sequence, we find the values of $v$, $\kappa_\alpha$, $\bm{G}$ and $\tilde{\bm{U}}_\alpha$. Inserting these values into Eq.~(\ref{eq:gen_sol}) gives the plane-wave solution of the SAW.
Note that the overall amplitude of $\bm{U}$ must be determined from the dimensions of IDTs and the applied rf power.
We describe this process in the Appendix.

\subsection{\label{sec:coordinate}Coordinate transformation}
Since the mechanical and electrical properties of the piezoelectric substrate (e.g. LiNbO$_3$) are anisotropic, characteristics of the SAW vary depending on $\theta_\mathrm{SAW}$, the relative angle between the propagation direction and the crystalline axis. We thus perform a coordinate transformation to take the anisotropy into account. 

For piezoelectric materials, the elastic, dielectric, and piezoelectric constants are often expressed with respect to the crystalline coordinate axis (the crystalline frame). 
The coordinate must be transformed if the lab frame does not match that of the crystal.
First, we express $T_{ij}$ and $S_{ij}$ using the Voigt representation:
\begin{equation}
	\begin{aligned}
		\bm{T} &\equiv [T_1 \quad T_2 \quad T_3 \quad T_4 \quad T_5 \quad T_6]^{T} \\
		&= [T_{xx} \quad T_{yy} \quad T_{zz} \quad T_{yz} \quad T_{zx} \quad T_{xy} ]^{T}\\
		\bm{S} &\equiv [S_1 \quad S_2 \quad S_3 \quad S_4 \quad S_5 \quad S_6]^{T}\\
		& = [S_{xx} \quad S_{yy} \quad S_{zz} \quad 2S_{yz} \quad 2S_{zx} \quad 2S_{xy}]^{T}.
		\label{eq:voigt}
	\end{aligned}
\end{equation}
Employing the Voigt representation, the piezoelectric relations described in Eqs.~(\ref{eq:piezo_eq_fund:T}) and (\ref{eq:piezo_eq_fund:D}) are rewritten as:
\begin{equation}
	\begin{aligned}
		\bm{T} &= \mathcal{C} \bm{\epsilon} - e^T \bm{E},\\
		\bm{D} &= e \bm{S} + \mathcal{E} \bm{E},
		\label{eq:piezo_eq}
	\end{aligned}
\end{equation}
where $\mathcal{C}$, $e$, and $\mathcal{E}$ are $6\times 6$, $3 \times 6$, and $3\times 3$ matrices, respectively.

The following rotation matrices are defined to transform the coordinate:
$\mathcal{R}$ is the standard $3 \times 3$ rotation matrix that rotates the coordinate from the crystalline frame to the lab frame.
$\mathcal{L}$ is a matrix that transforms the stress and the strain~\cite{campbell1968ieee}. 
They are defined as
\begin{equation}
	\begin{aligned}
		&\mathcal{R}\equiv\mqty[
		R_{xx} & R_{xy} & R_{xz} \\
		R_{yx} & R_{yy} & R_{yz} \\
		R_{zx} & R_{zy} & R_{zz} \\
		].
		\label{eq:rot_R}
	\end{aligned}
\end{equation}
\begin{widetext}
\begin{equation}
	\begin{aligned}
		&\mathcal{L}=\mqty[
		R_{xx}^2 & R_{xy}^2 & R_{xz}^2 & 2R_{xy}R_{xz} & 2R_{xz}R_{xx} & 2R_{xx}R_{xy} \\
		R_{yx}^2 & R_{yy}^2 & R_{yz}^2 & 2R_{yy}R_{yz} & 2R_{yz}R_{yx} & 2R_{yx}R_{yy} \\
		R_{zx}^2 & R_{zy}^2 & R_{zz}^2 & 2R_{zy}R_{zz} & 2R_{zz}R_{zy} & 2R_{zx}R_{zy} \\
		R_{yx}R_{zx} & R_{yy}R_{zy} & R_{yz}R_{zz} & R_{zy}R_{yz}+R_{zz}R_{yy} & R_{zx}R_{xz}+R_{zz}R_{xx} & R_{zx}R_{yy}+R_{zy}R_{yx}  \\
		R_{zx}R_{xx} & R_{zy}R_{xy} & R_{zz}R_{xz} & R_{zy}R_{xz}+R_{zz}R_{xy} & R_{zx}R_{xz}+R_{zz}R_{xx} & R_{zx}R_{xy}+R_{zy}R_{xx}  \\
		R_{xx}R_{yx} & R_{xy}R_{yy} & R_{xz}R_{yz} & R_{yy}R_{xz}+R_{yz}R_{xy} & R_{yx}R_{xz}+R_{yz}R_{xx} & R_{yx}R_{xy}+R_{yy}R_{xx}
		],
		\label{eq:rot:LT}
	\end{aligned}
\end{equation}
\end{widetext}
Upon transforming the coordinate from the crystalline frame to the lab frame, the piezoelectric relations read
\begin{equation}
	\begin{aligned}
		\bm{T'}&= \mathcal{C}' \bm{\epsilon'} - e'^T \bm{E'}, \\
		\bm{D'}&= e' \bm{\epsilon'} + \mathcal{E}' \bm{E'},
		\label{eq:piezo_eq_trans}
	\end{aligned}
\end{equation}
where 
\begin{equation}
	\begin{aligned}
		\bm{T'} = \mathcal{L} \bm{T}, \ \ \bm{\epsilon'} = \mathcal{L}^T \bm{\epsilon}, \ \ \bm{D'} = \mathcal{R} \bm{D}, \ \ \bm{E'} = \mathcal{R} \bm{E},
		\label{eq:piezo_eq_trans_quants}
	\end{aligned}
\end{equation}
\begin{equation}
	\begin{aligned}
		\mathcal{C}' = \mathcal{L} \mathcal{C} \mathcal{L}^T, \ \ e' = \mathcal{R} e \mathcal{L}^T,\ \ \mathcal{E}' = \mathcal{R} \mathcal{E} \mathcal{R}^T.
		\label{eq:piezo_eq_trans_consts}
	\end{aligned}
\end{equation}

\subsection{Material constants for LiNbO$_3$}

Single crystalline lithium niobate (LiNbO$_3$) is often chosen as a piezoelectric substrate since it has a large electromechanical coupling coefficient.
Its crystal structure belongs to the space group $R3c$ and lacks inversion symmetry.
LiNbO$_3$ has a spontaneous polarization, which causes the piezoelectricity.
The crystalline $Z$-axis is taken along the spontaneous polarization and the $X$-axis is chosen along the orientation connecting the Nb atoms. 
Material constants of LiNbO$_3$ are expressed in the crystalline frame ($XYZ$).
Reflecting the crystal symmetry, $\mathcal{C}$, $e$, and $\mathcal{E}$ of LiNbO$_3$ are expressed in the following forms~\cite{warner1967jasa}:
\begin{equation}
	\begin{gathered}
		\mathcal{C}=\mqty[
		c_{11} & c_{12} & c_{13} & c_{14} & 0 & 0 \\
		c_{12} & c_{11} & c_{13} & -c_{14} & 0 & 0 \\
		c_{13} & c_{13} & c_{33} & 0 & 0 & 0 \\
		c_{14} & -c_{14} & 0 & c_{44} & 0 & 0 \\
		0 & 0 & 0 & 0 & c_{44} & c_{14} \\
		0 & 0 & 0 & 0 & c_{14} & c_{66}
		],\\
		e=\mqty[
		0 & 0 & 0 & 0 & e_{15} & -e_{22}\\
		-e_{22} & e_{22} & 0 & e_{15} & 0 & 0 \\
		e_{31} & e_{31} & e_{33} & 0 & 0 & 0
		],
		\ \ 
		\mathcal{E}=\mqty[
		\varepsilon_{11} & 0 & 0 \\
		0 & \varepsilon_{11} & 0 \\
		0 & 0 & \varepsilon_{33}
		].
		\label{eq:piezo_const}
	\end{gathered}
\end{equation}
Here we used the Voigt representation. The parameters of LiNbO$_3$ are summarized in Tab.~\ref{tab:unit_cell_LN}-\ref{tab:die_consts_ln}~\cite{wong2002book,kushibiki1999ieee}.
\begin{table}[htb]
	\centering
	\begin{tabular}{|c|c|c|}
		\hline
		$a$ [\AA] & $c$ [\AA] & $\rho_\mathrm{p}$ [kg/m$^3$] \\ \hline \hline
		5.15 & 13.9 & 4642.8 \\ \hline 
	\end{tabular}
	\caption{
		Unit cell parameters~\cite{wong2002book} and the mass density~\cite{kushibiki1999ieee} of LiNbO$_3$.
	}
	\label{tab:unit_cell_LN}
\end{table}

\begin{table}[hbt]
	\centering
	\begin{tabular}{|c||c|}
		\hline
		Coefficient & Value [$10^{11}$ N/m$^2$]  \\ \hline \hline
		$c_{11}$ & 1.9886 \\ \hline 
		$c_{12}$ & 0.5467 \\ \hline 
		$c_{13}$ & 0.6799 \\ \hline 
		$c_{14}$ & 0.0783 \\ \hline 
		$c_{33}$ & 2.3418 \\ \hline 
		$c_{44}$ & 0.5985 \\ \hline 
		$c_{66}$ & 0.7209 \\ \hline
	\end{tabular}
	\caption{
		Elastic constants of LiNbO$_3$~\cite{kushibiki1999ieee}.
	}
	\label{tab:elastic_consts_ln}
\end{table}

\begin{table}[hbt]
	\centering
	\begin{tabular}{|c||c|}
		\hline
		Coefficient & Value [C/m$^2$]  \\ \hline \hline
		$e_{15}$ & 3.655 \\ \hline 
		$e_{22}$ & 2.407 \\ \hline 
		$e_{31}$ & 0.328 \\ \hline 
		$e_{33}$ & 1.894 \\ \hline 
	\end{tabular}
	\caption{
		Piezoelectric constants of LiNbO$_3$~\cite{kushibiki1999ieee}.
	}
	\label{tab:piezo_consts_ln}
\end{table}

\begin{table}[hbt]
	\centering
	\begin{tabular}{|c||c|}
		\hline
		Coefficient & Value [arb.]  \\ \hline \hline
		$\varepsilon_{11}/\varepsilon_0$ & 44.9 \\ \hline 
		$\varepsilon_{33}/\varepsilon_0$ & 26.7 \\ \hline 
	\end{tabular}
	\caption{
		Dielectric constants of LiNbO$_3$ under constant strains~\cite{kushibiki1999ieee}.
		$\varepsilon_0$ is the vacuum permittivity.
	}
	\label{tab:die_consts_ln}
\end{table}

\begin{figure}[b]
	\centering
	\begin{minipage}{1.0\hsize}
		\centering
		\includegraphics[scale=0.07]{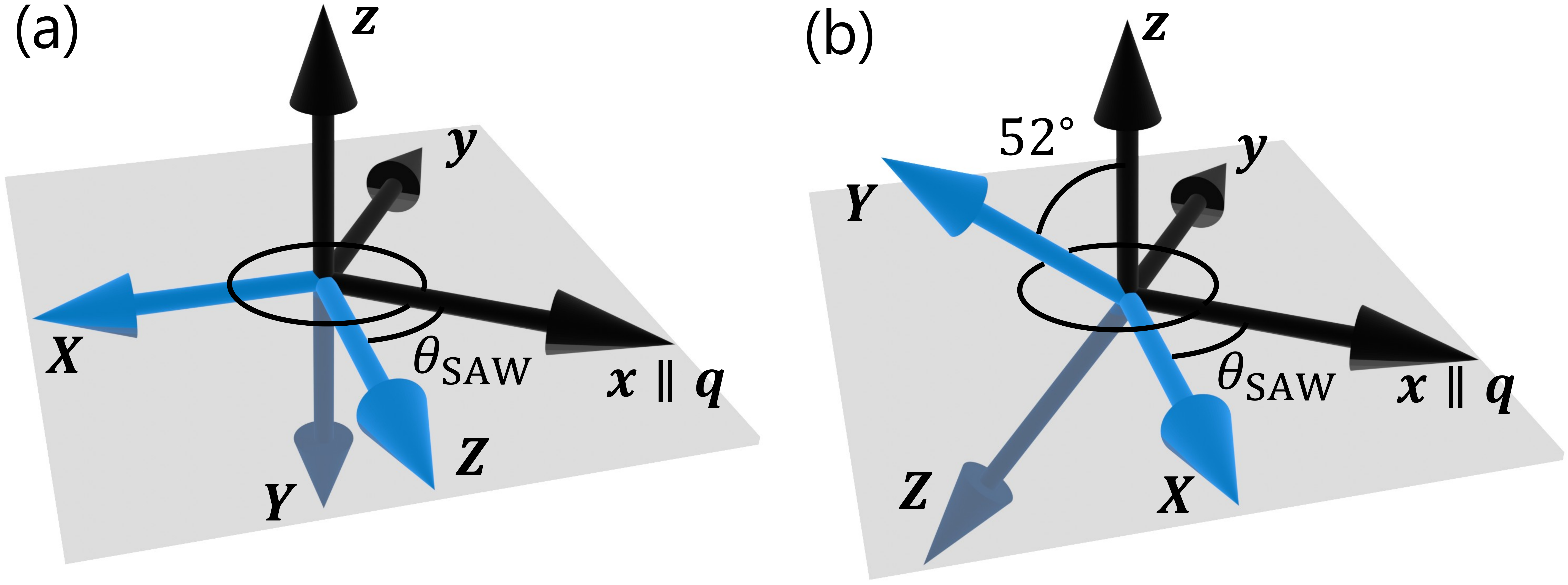}
	\end{minipage}
	\caption{
		(a,b) Crystalline and lab coordinates of Y-cut (a) and Y$+128^\circ$-cut (b) LiNbO$_3$ substrates. Blue ($X,Y,Z$) arrows are the crystalline coordinate of LiNbO$_3$. 
		$Z$-axis corresponds to the direction of the spontaneous polarization. The black ($x,y,z$) arrows show the lab coordinate axes. $\theta_\mathrm{SAW}$ is the relative angle between the SAW propagation direction $\bm{q}\parallel x$ and the $Z$-axis (a) or the $X$-axis (b).
		\label{fig:Fig2}
	}
\end{figure}

Y-cut and Y$+128^\circ$-cut LiNbO$_3$ substates are often used for SAW devices, which are illustrated in Fig.~\ref{fig:Fig2}.
The surface normal of a Y-cut LiNbO$_3$ is parallel to the $Y$-axis whereas the surface normal of a Y$+128^\circ$-cut LiNbO$_3$ is parallel to an orientation where the $Y$-axis is rotated $128^\circ$ around the $X$-axis, corresponding to the $\langle 10\bar{1}4 \rangle$ crystallographic orientation~\cite{polewczyk2019iop}.
It is known that SAW is most efficiently excited when the SAW propagation direction is parallel to the $Z$-axis ($X$-axis) for the Y-cut (Y$+128^\circ$-cut) LiNbO$_3$. 
We therefore define the angle between the SAW wave vector and the $Z$-axis ($X$-axis) of the Y-cut (Y$+128^\circ$-cut) LiNbO$_3$ as $\theta_\mathrm{SAW}$.
To transform the coordinate from the crystalline frame ($XYZ$) to the lab frame ($xyz$) shown in Fig.~\ref{fig:Fig2}, the rotation matrix $\mathcal{R}$ for Y-cut LiNbO$_3$ is expressed by
\begin{equation}
	\begin{aligned}
		&\mathcal{R}\equiv\mqty[
		-\sin(\theta_\mathrm{SAW}) & 0 &  \cos(\theta_\mathrm{SAW}) \\
		-\cos(\theta_\mathrm{SAW}) & 0 & -\sin(\theta_\mathrm{SAW}) \\
		0 & -1 & 0 \\
		],
		\label{eq:rot_R_Ycut}
	\end{aligned}
\end{equation}
and that for Y$+128^\circ$-cut LiNbO$_3$ is
\small
\begin{equation}
	\begin{aligned}
		&\mathcal{R}\equiv\mqty[
		\cos(\theta_\mathrm{SAW}) & -\cos{38^\circ}\sin(\theta_\mathrm{SAW}) & -\sin{38^\circ}\sin(\theta_\mathrm{SAW}) \\
		-\sin(\theta_\mathrm{SAW}) & -\cos{38^\circ}\cos(\theta_\mathrm{SAW}) & -\sin{38^\circ}\cos(\theta_\mathrm{SAW}) \\
		0 & \sin{38^\circ} & -\cos{38^\circ} \\
		].
		\label{eq:rot_R_128Ycut}
	\end{aligned}
\end{equation}
\normalsize
Representative parameters of the SAW are its phase velocity and the electromechanical coupling constant $K_\mathrm{eff}^2$ (see also Eq.~(\ref{eq:keff2_def}) in the Appendix). Here we denote the SAW velocity under the open and shorted boundary conditions as $v_0$, $v_m$, respectively. Values of $v_0$, $v_m$, and $K_\mathrm{eff}^2$ for LiNbO$_3$ from the literature~\cite{datta1986book} are summarized in Table~\ref{tab:0deg_v_keff}.
\begin{table}[hbt]
	\centering
	\begin{tabular}{|c||c|c|c|c|}
		\hline
		Cut angle & Propagation axis &$v_0$ [m/s] & $v_m$ [m/s] & $K^2_\mathrm{eff}$ [\%] \\ \hline \hline
		Y & $Z$ & 3488 & 3408 & 4.6 \\ \hline 
		Y$+128^\circ$ & $X$ & 3996 & 3884 & 5.6 \\ \hline
	\end{tabular}
	\caption{
		SAW velocity and the electromechanical coupling coefficient~\cite{datta1986book}.
		The substrate represents the cut angle of LiNbO$_3$.
		The axis represents the SAW propagation orientation.
	}
	\label{tab:0deg_v_keff}
\end{table}

\section{Results}

\begin{figure}[t]
	\centering
	\begin{minipage}{1.0\hsize}
		\centering
		\includegraphics[scale=0.25]{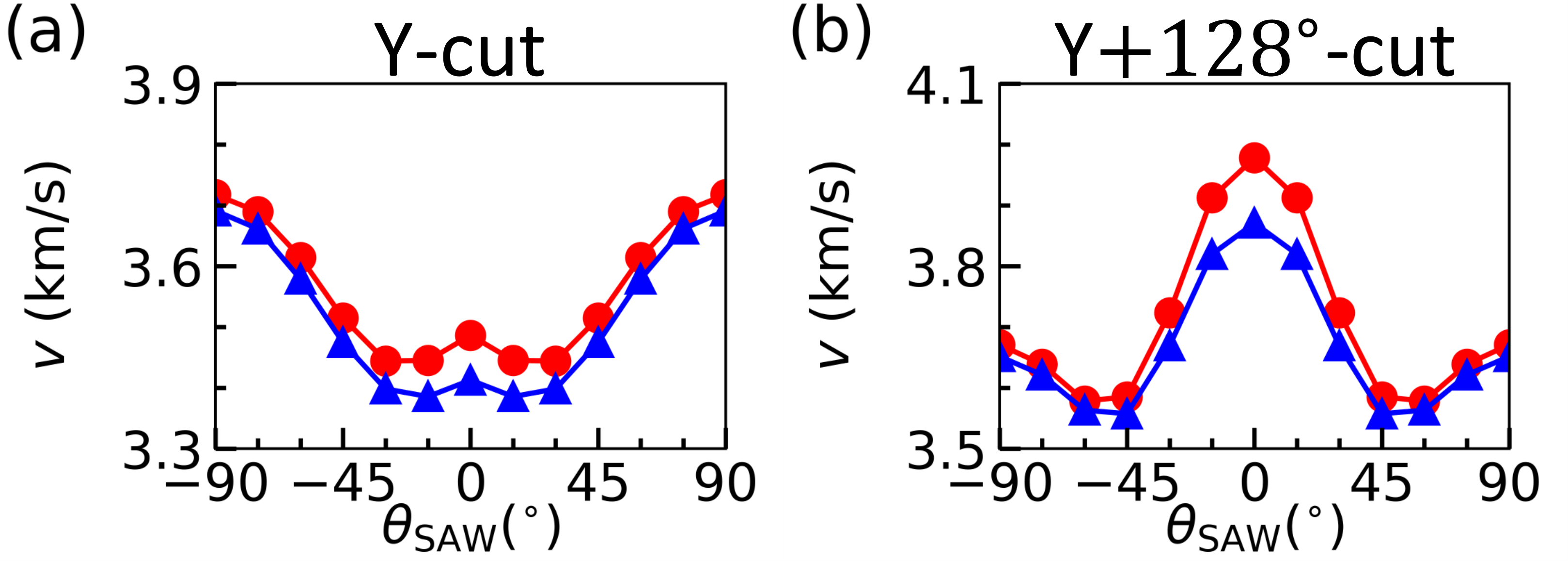}
	\end{minipage}
	\caption{
		(a,b) SAW propagation direction ($\theta_\mathrm{SAW}$) dependence of the SAW velocity $v$ for Y-cut (a) and Y$+128^\circ$-cut (b) LiNbO$_3$. Red circles (blue triangles) represent the results under the open (shorted) boundary condition.
		\label{fig:Fig3_rev}
	}
\end{figure}

In this section, we present numerically calculated propagation characteristics of SAWs as a function of $\theta_\mathrm{SAW}$. 
The variation of the phase velocity $v$ with $\theta_\mathrm{SAW}$ is presented in Fig.~\ref{fig:Fig3_rev}. 
Red and blue symbols indicate values that are obtained under open and shorted boundary conditions. 
The smaller $v$ obtained under the shorted boundary condition (blue symbols) is caused by the change in the electric field distribution at the substrate/film interface from that of the substrate/vacuum interface~\cite{datta1986book}.

\begin{figure}[t]
	\centering
	\begin{minipage}{1.0\hsize}
		\centering
		\includegraphics[scale=0.25]{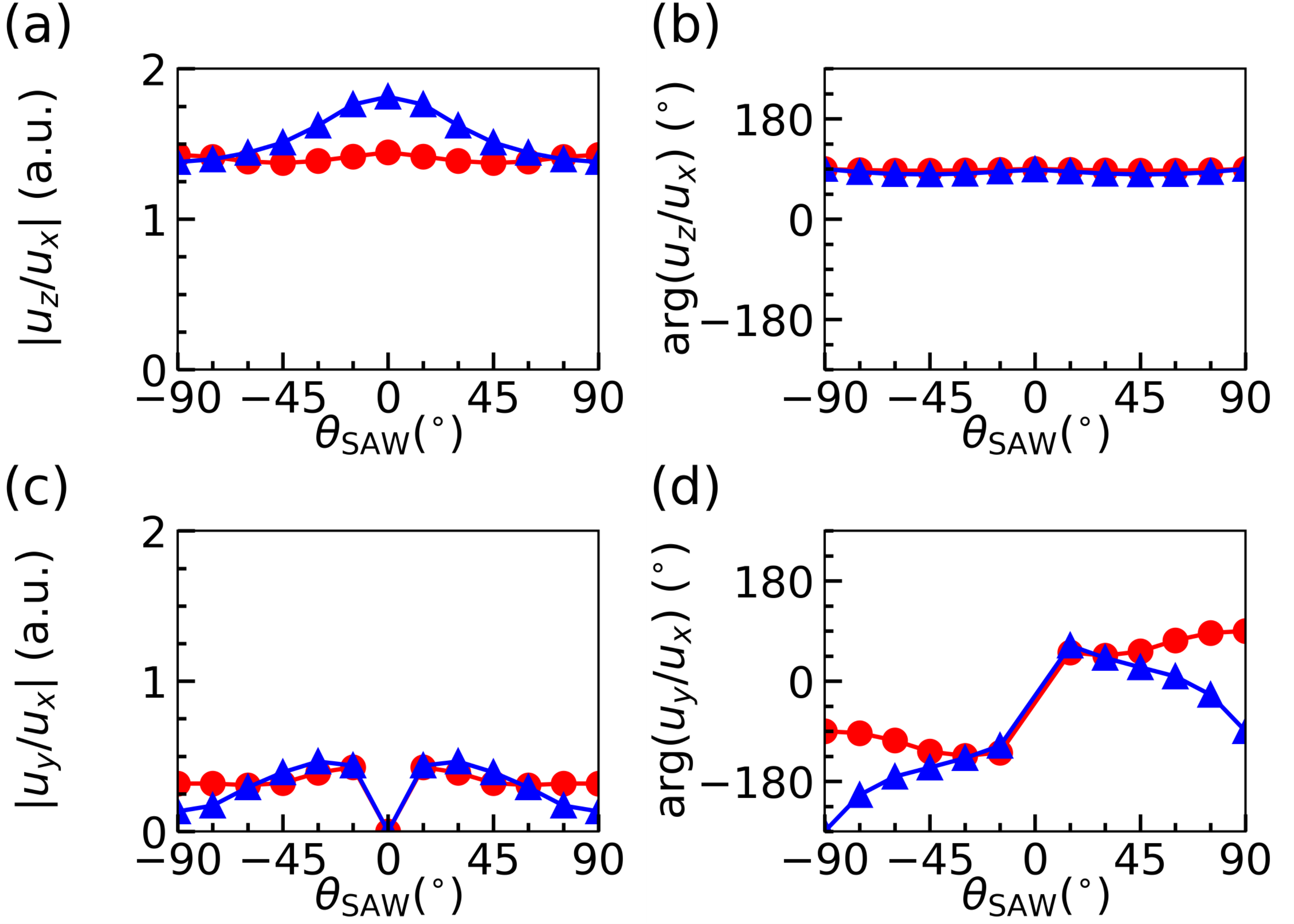}
	\end{minipage}
	\caption{
		(a-d) SAW propagation direction ($\theta_\mathrm{SAW}$) dependence of the SAW lattice displacements for Y-cut LiNbO$_3$. The graphs show $\abs{u_z/u_x}$ (a), $\mathrm{arg}\qty(u_z/u_x)$ (b), $\abs{u_y/u_x}$ (c), and $\mathrm{arg}\qty(u_y/u_x)$ (d), respectively. Red circles (blue triangles) represent the results under the open (shorted) boundary condition. The values of $u_z/u_x,u_y/u_x$ are those at the substrate surface ($z=0$). 
		$\mathrm{arg}\qty(u_y/u_x)$ for $\theta_\mathrm{SAW}=0^\circ$ is not plotted in (d) since the corresponding amplitude is zero.
		\label{fig:Fig4_rev}
	}
\end{figure}

\begin{figure}[t]
	\centering
	\includegraphics[scale=0.2]{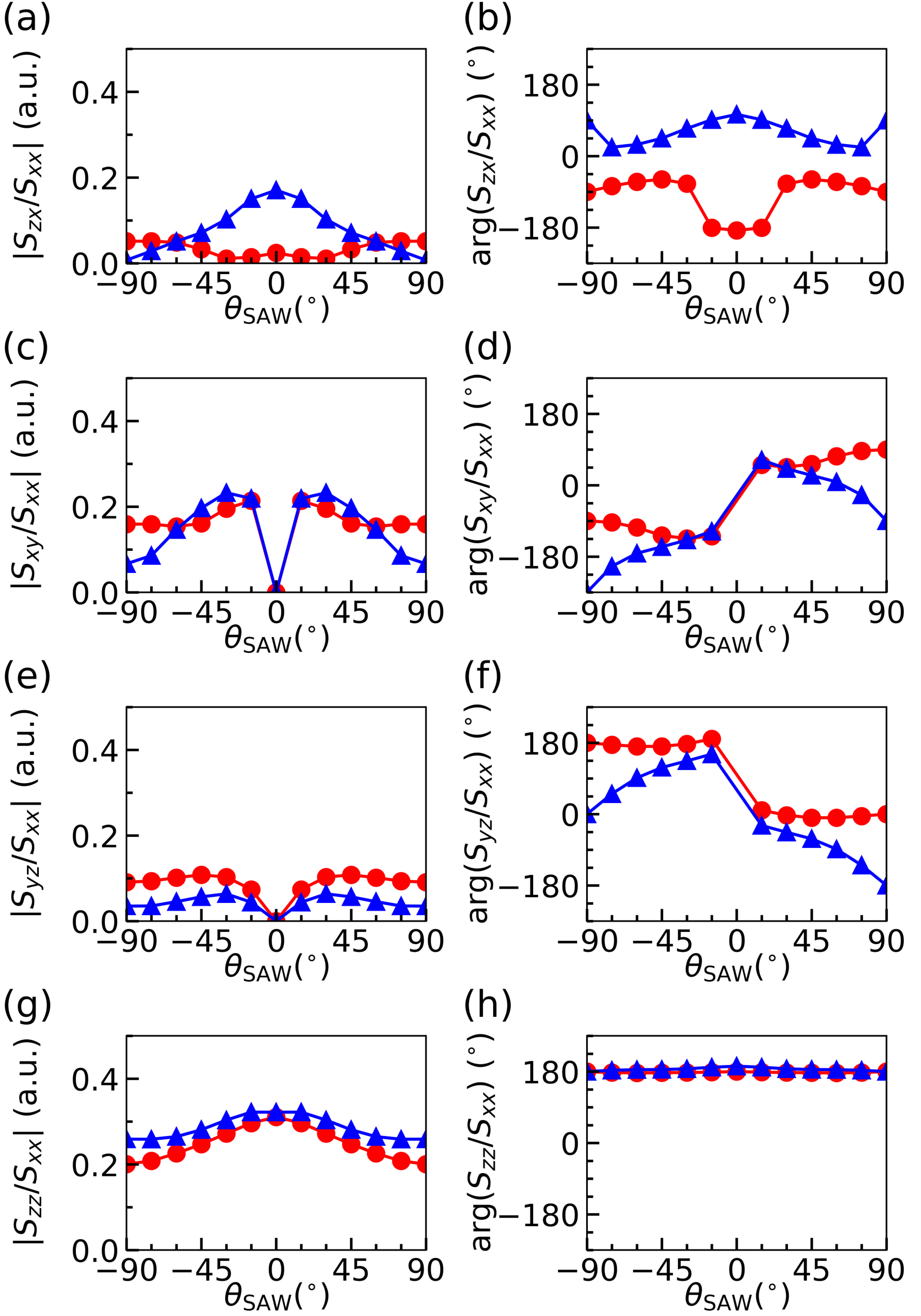}
	\caption{
		(a-h) SAW propagation direction ($\theta_\mathrm{SAW}$) dependence of the relative amplitudes (a,c,e,g) and phases (b,d,f,h) among the quantities associated with the SAW traveling in Y-cut LiNbO$_3$. The values at the substrate surface ($z=0$) are presented. The graphs show the results of $S_{zx}/S_{xx}$ (a,b), $S_{xy}/S_{xx}$ (c,d), $S_{yz}/S_{xx}$ (e,f), and $S_{zz}/S_{xx}$ (g,h), respectively. Red circles (blue triangles) represent the results under the open (shorted) boundary condition. 
		$\mathrm{arg}\qty(S_{xy}/S_{xx})$ and $\mathrm{arg}\qty(S_{yz}/S_{xx})$ for $\theta_\mathrm{SAW}=0^\circ$ are not plotted in (d,f) since the corresponding amplitudes are zero.
		\label{fig:Fig5_rev}
	}
\end{figure}

Figures~\ref{fig:Fig4_rev} and \ref{fig:Fig5_rev} show the calculated displacement and strain for SAWs excited on a Y-cut substrate. Results for those on a Y$+128^{\circ}$-cut substrate are presented in the Appendix.
In Fig.~\ref{fig:Fig4_rev}, the displacements $u_y$ and $u_z$ are normalized by $u_x$. Similarly, all strain components presented in Fig.~\ref{fig:Fig5_rev} are normalized by $S_{xx}$.
The amplitudes of $u_x$ and $S_{xx}$ can be estimated if the dimensions of IDTs, including the pitch of the IDT, the number of finger pairs $N_s$ and the IDT aperture $W$, and the rf power input to the IDTs are specified. See Figs.~\ref{fig:Fig9_rev}(g) and \ref{fig:Fig9_rev}(h) in the Appendix for an example of the evaluation of $S_{xx}$.

The relative phase of the displacements $u_z$ and $u_x$ is $\approx 90^\circ$ (see Fig.~\ref{fig:Fig4_rev}(b)), indicating that the obtained SAW mode is the Rayleigh-type. 
Figure~\ref{fig:Fig4_rev}(c) shows that a non-negligible displacement along $y$ ($u_y$) and, consequently, shear strains $S_{xy}$ and $S_{yz}$ [Figs.~\ref{fig:Fig5_rev}(c) and (e)] emerge, which are absent in the Rayleigh-type SAW excited in nonpiezoelectric homogeneous media. 
With regard to magnon-phonon coupling, the magnetoelastic energy $E_\mathrm{me}$ that mediates the coupling is often expressed in the following form:
\begin{equation}
	\begin{aligned}
		E_\mathrm{me} &= \frac{b_1}{M_\mathrm{s}} \left( S_{xx} m_x^2 + S_{yy} m_y^2 + S_{zz} m_z^2 \right)\\
		& + \frac{2b_2}{M_\mathrm{s}} \left( S_{xy} m_x m_y + S_{yz} m_y m_z + S_{zx} m_z m_x \right),
		\label{eq:energy:me}
	\end{aligned}
\end{equation}
where $b_1$ and $b_2$ are the magnetoelastic energy, $M_\mathrm{s}$ is the saturation magnetization, and $\bm{m}=(m_x,m_y,m_z)$ represents the direction of magnetization. 
The results presented in Fig.~\ref{fig:Fig5_rev} indicate that the magnetoelastic coupling arising from the shear strains cannot be neglected for SAWs traveling in certain directions. The coupling induced by such strain components can be detected using, for example, the $\theta_\mathrm{SAW}$ dependence of the SAW power absorption due to excitation of spin waves~\cite{dreher2012prb}.

The electrical boundary conditions influence both the relative amplitude and phase among the strain components. In particular, $S_{zx}$ is suppressed under the open boundary condition for the Y-cut substrate. We remark that the opposite trend is observed for the Y$+128^\circ$-cut substrate as shown in Fig.~\ref{fig:Fig7_rev}(a) in the Appendix.
However, it should be noted that our calculations assume a mechanically free surface. In practice, the presence of a mechanical load (e.g. a thick film) on the path where the SAW travels may lead to quantitatively different behaviors~\cite{tateno2020prapp}, which remains a subject for a further study.


Experimentally, the conductive film deposited on substrate possesses finite thickness and conductivity. 
In such case, strictly speaking, the shorted boundary condition cannot be applied. This issue can be addressed by introducing a specific surface impedance $Z_\mathrm{p}$ for the film with given thickness and conductivity~\cite{ingebrigtsen1970jap,kawada2024arXiv}.
Although a detailed discussion on this subject is beyond the scope of this work, it is worth noting that, at least for thin metallic films, all strain components take a value similar to those obtained under the shorted boundary condition.

\section{Summary}
In summary, we present a comprehensive methodology for numerically analyzing the propagation characteristics of surface acoustic waves (SAWs) excited on a piezoelectric medium. As an example, we show computation results for Rayleigh-type SAWs propagating in LiNbO$_3$ substrates. 
We find that the electrical boundary conditions at the LiNbO$_3$ surface, whether it faces vacuum or is covered with a perfect conductor, influence the SAW propagation properties, such as the phase velocity, the amplitude and phase of the strain tensor components. 
Due to the crystalline anisotropy of the substrate, all these quantities exhibit variation with the SAW propagation direction. Notably, in contrast to non-piezoelectric isotropic media, we demonstrate that all shear strain components can be induced in LiNbO$_3$.   
These results are useful for studying the SAW-driven magnon-phonon coupling mediated by magnetoelastic and spin-rotation couplings and provide valuable insights for engineering strain to develop functional magnonic and phononic devices.

\section{Appendix}
\subsection{Results for Y$+128^\circ$-cut LiNbO$_3$ substrate}

\begin{figure}[b]
	\centering
	\begin{minipage}{1.0\hsize}
		\centering
		\includegraphics[scale=0.25]{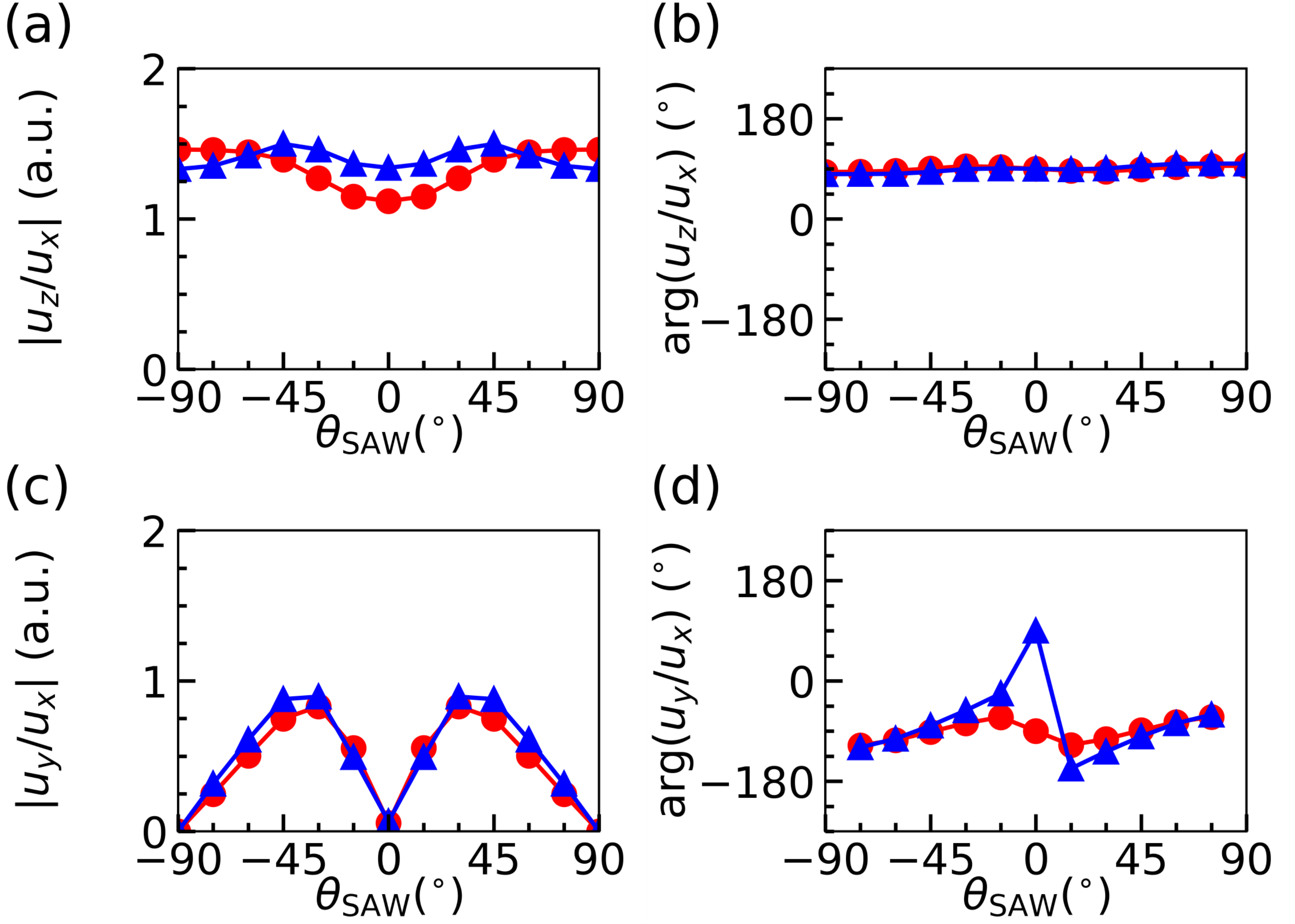}
	\end{minipage}
	\caption{
		(a-d) SAW propagation direction ($\theta_\mathrm{SAW}$) dependence of the SAW lattice displacements for Y$+128^\circ$-cut LiNbO$_3$. The graphs show $\abs{u_z/u_x}$ (a), $\mathrm{arg}\qty(u_z/u_x)$ (b), $\abs{u_y/u_x}$ (c), and $\mathrm{arg}\qty(u_y/u_x)$ (d), respectively. Red circles (blue triangles) represent the results under the open (shorted) boundary condition. The values of $u_z/u_x,u_y/u_x$ are those at the substrate surface ($z=0$). 
		$\mathrm{arg}\qty(u_y/u_x)$ for $\theta_\mathrm{SAW}=\pm 90^\circ$ is not plotted in (d) since the corresponding amplitude is zero.
		\label{fig:Fig6_rev}
	}
\end{figure}

Here we show the propagation characteristics of SAWs computed for Y$+128^\circ$-cut LiNbO$_3$ as a function of $\theta_\mathrm{SAW}$. 
The obtained SAW mode is indicated to be the Rayleigh-type, since the relative phase of the displacements $u_z$ and $u_x$ is $\approx 90^\circ$ (see Fig.~\ref{fig:Fig6_rev}(b)).

\begin{figure}[b]
	\centering
	\includegraphics[scale=0.2]{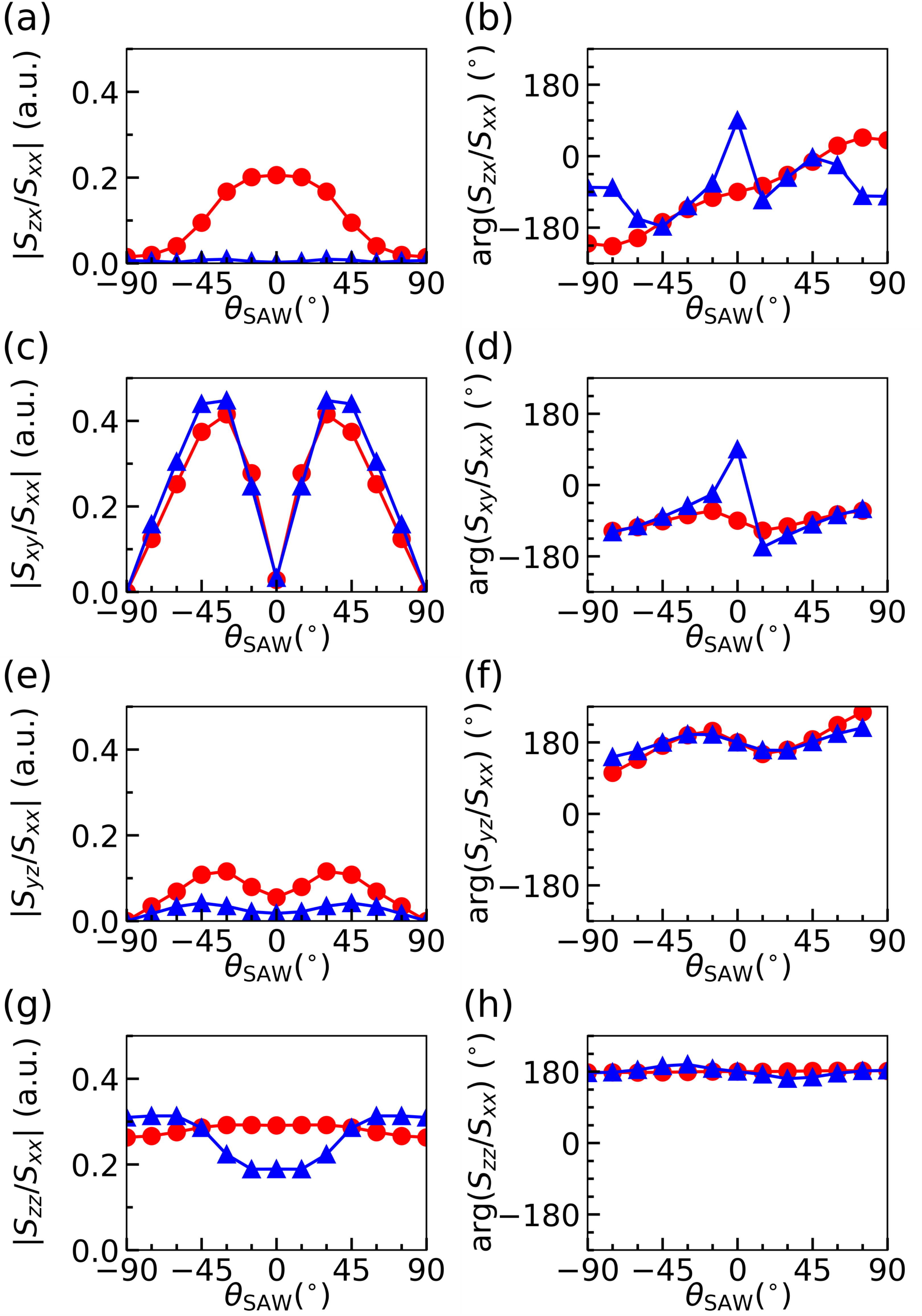}
	\caption{
		(a-h) SAW propagation direction ($\theta_\mathrm{SAW}$) dependence of the relative amplitudes (a,c,e,g) and phases (b,d,f,h) among the quantities associated with the SAW traveling in Y$+128^\circ$-cut LiNbO$_3$. The values at the substrate surface ($z=0$) are presented. The graphs show the results of $S_{zx}/S_{xx}$ (a,b), $S_{xy}/S_{xx}$ (c,d), $S_{yz}/S_{xx}$ (e,f), and $S_{zz}/S_{xx}$ (g,h), respectively. Red circles (blue triangles) represent the results under the open (shorted) boundary condition. 
		$\mathrm{arg}\qty(S_{xy}/S_{xx})$ and $\mathrm{arg}\qty(S_{yz}/S_{xx})$ for $\theta_\mathrm{SAW}=\pm 90^\circ$ are not plotted in (d,f) since the corresponding amplitudes are zero.
		\label{fig:Fig7_rev}
	}
\end{figure}

The SAW in Y$+128^\circ$-cut LiNbO$_3$ exhibits some quantitative differences from that in Y-cut: in Fig.~\ref{fig:Fig7_rev}(a), $S_{zx}$ is significantly suppressed under the shorted boundary condition, whereas it is enhanced for the Y-cut LiNbO$_3$ with the shorted surface (see Fig.~\ref{fig:Fig5_rev}(a)). The SAW in Y$+128^\circ$-cut LiNbO$_3$ also contains a non-negligible displacement along $y$ ($u_y$) and thus the shear strains $S_{xy}$ and $S_{yz}$ [Figs.~\ref{fig:Fig7_rev}(c) and \ref{fig:Fig7_rev}(e)]. It is noteworthy that the maximum amplitude of $S_{xy}/S_{xx}$ for Y$+128^\circ$-cut LiNbO$_3$ is twice as large as that for Y-cut, which might be benificial to study, for instance, the magnetoacoustic resonance caused by the out-of-plane vorticity~\cite{mingxian2023} or phononic angular momenta~\cite{liyang2024}.

\subsection{\label{sec:supp:calcsawamp}Quantitative estimation of SAW amplitude}

In the experiments, SAWs are launched at the IDT by applying an rf signal.
The SAWs travel along the substrate surface, passing a region that faces vacuum (or air) and reaches the region of interest, where a film is placed.
Here we assume the film is an infinitesimally thin perfect conductor.
To study the SAW characteristics at the region of interest, we must determine (1) the SAW excitation efficiency at the IDT, i.e. the efficiency to convert the applied rf signal to the SAW amplitude at the IDT, (2) the SAW amplitude that travels the substrate/vacuum interface and (3) the SAW amplitude that reaches the region of interest.

\begin{figure}[b]
	\centering
	\begin{minipage}{1.0\hsize}
		\centering
		\includegraphics[scale=0.075]{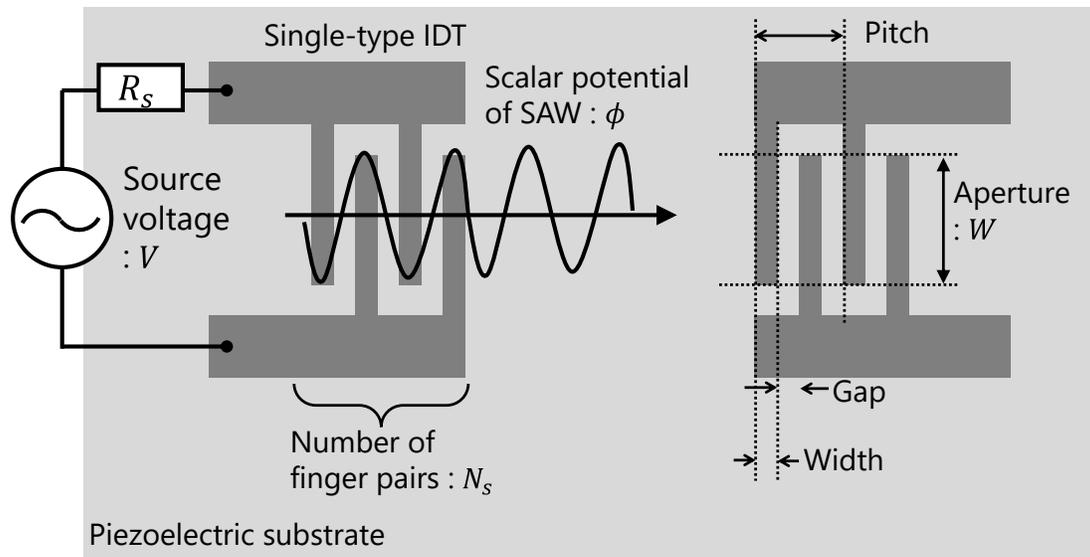}
	\end{minipage}
	\caption{
		Schematic illustration of a single-type IDT.
		\label{fig:Fig8_rev}
	}
\end{figure}

\begin{figure}[b]
	\centering
	\includegraphics[scale=0.2]{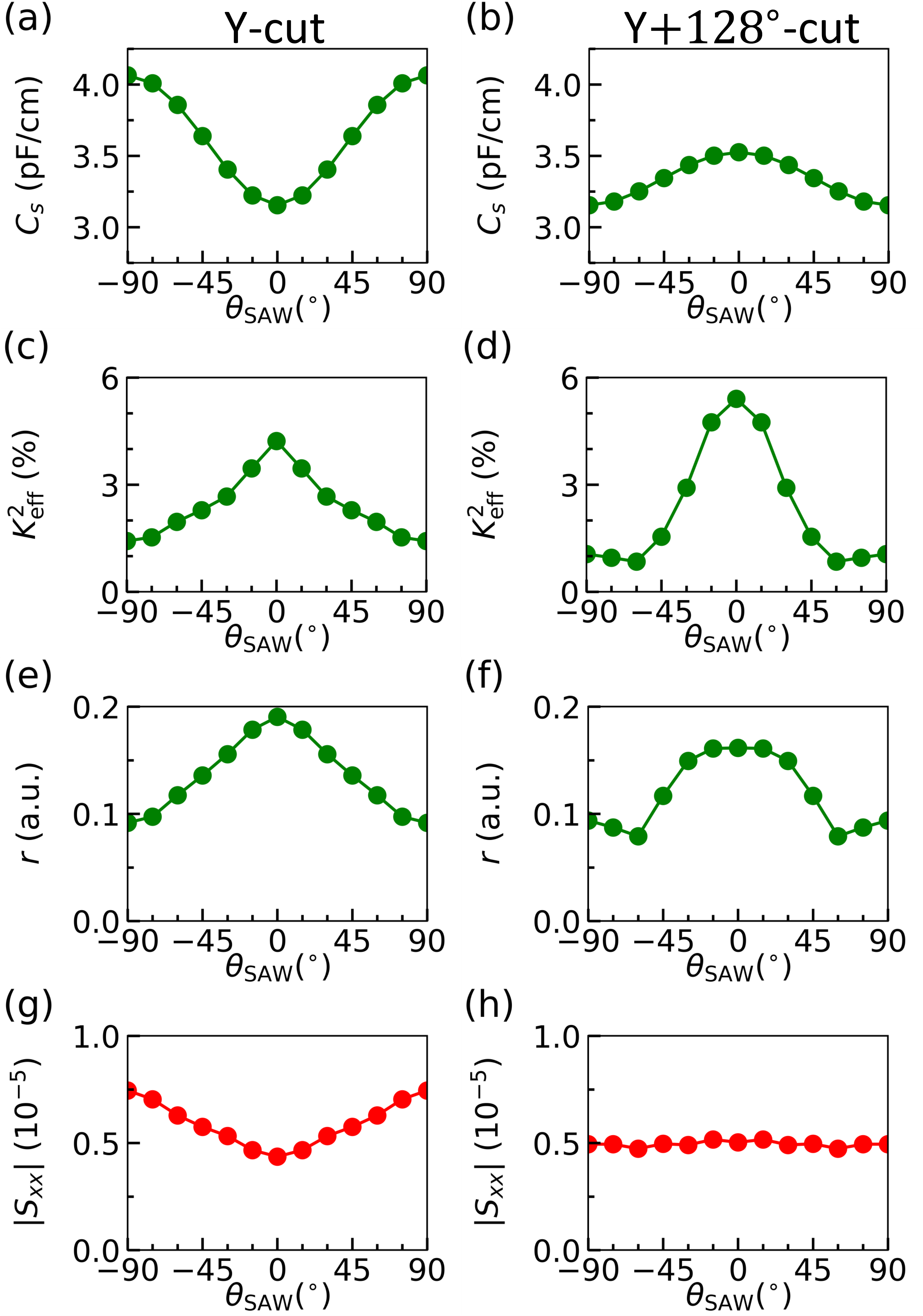}
	\caption{
		(a-h) SAW propagation direction ($\theta_\mathrm{SAW}$) dependence of the SAW characteristics for Y-cut (a,c,e,g) and Y$+128^\circ$-cut (b,d,f,h) LiNbO$_3$. The graphs show $C_s$ (a,b), $K_\mathrm{eff}^2$ (c,d), $r$ (e,f), and $|S_{xx}|$ (g,h), respectively. $r$ is estimated under the condition of $W=370\ \upmu$m and $N=40$ using Eq.~(\ref{eq:Piezo_potential}). The values of $S_{xx}$ are those at the substrate surface ($z=0$) and $V=0.32$ V. 
		\label{fig:Fig9_rev}
	}
\end{figure}

We start from the SAW excitation efficiency $r$, which is defined as the ratio of the scalar potential $\phi$ associated with the SAW to the rf source voltage $V$.
Note that $r$ is dependent on the rf source angular frequency $\omega$ and the geometry of the IDT.
Here we assume a single type IDT where the signal and ground fingers are alternately aligned, as schematically shown in Fig.~\ref{fig:Fig8_rev}.
The ratio of the finger width to pitch, referred to as the metallization, is set to 0.5.
For simplicity, we assume that $\omega$ is equal to the center value of the SAW resonant frequency.
The expression of $r$ for such IDT is described as~\cite{datta1986book}:
\begin{equation}
	\begin{aligned}
		r \equiv \abs{\frac{\phi}{V}} = \sqrt{\frac{T}{2\tau}\cdot \frac{N_s K_\mathrm{eff}^2 }{\qty(1 + \omega T)^2 +\qty(\omega \tau)^2}},
		\label{eq:Piezo_potential}
	\end{aligned}
\end{equation}
with
\begin{equation}
	\begin{aligned}
		T\equiv 1.28 R_s C_s K_\mathrm{eff}^2 (N_s + 1)^2 W, \ \ \tau \equiv R_s C_s N_s W.
		\label{eq:x0_tau}
	\end{aligned}
\end{equation}
$R_s$ is the output impedance of the rf source, which is typically 50 $\Omega$, $W$ is the IDT aperture, $N_s$ is the number of finger pairs, and $K_\mathrm{eff}^2$ is the electromechanical coefficient defined as~\cite{ingebrigtsen1969jap,datta1986book}:
\begin{equation}
	\begin{aligned}
		K_\mathrm{eff}^2=2\abs{\frac{v_0 - v_m}{v_0}},
		\label{eq:keff2_def}
	\end{aligned}
\end{equation}
where $v_0$ and $v_m$ are the SAW velocity under the open and shorted boundary conditions, respectively. 
$C_s$ is the effective permittivity given by the following equation~\cite{datta1986book}:
\begin{equation}
	\begin{aligned}
		C_s = \varepsilon_0 + \sqrt{\varepsilon'_{11}\varepsilon'_{33} -\qty(\varepsilon'_{31})^2}.
		\label{eq:Cs_from_vareps}
	\end{aligned}
\end{equation}
$\varepsilon'_{ij}$ is the component of the dielectric tensor of the piezoelectric substrate defined in the lab coordinate.
$r$ depends on $\theta_\mathrm{SAW}$ due to the variation of $C_s$ and $K_\mathrm{eff}^2$ with $\theta_\mathrm{SAW}$.
The $\theta_\mathrm{SAW}$ dependence of $C_s$ and $K_\mathrm{eff}^2$ for LiNbO$_3$ are shown in Figs.~\ref{fig:Fig9_rev}(a)-\ref{fig:Fig9_rev}(d). $r$ can be estimated once $W$, $N_s$, and the pitch are given. As an example, we show the $\theta_\mathrm{SAW}$ dependence of $r$ in Figs.~\ref{fig:Fig9_rev}(e) and \ref{fig:Fig9_rev}(f) for $W=370\ \upmu$m, $N=40$, and a pitch of 2 $\upmu$m~\cite{tateno2020prapp}. 


Next we consider the amplitude of the SAW at the substrate/vacuum interface ($z=0$).
The calculation described in Sec.~II determines the plane-wave solution of the SAW except for its amplitude.
Let $\bm{U}'$ be one of such solutions.
From the discussion above, the amplitude of the SAW's scalar potential, or the forth component of $\bm{U}$ (see Eq.~(\ref{eq:plane_abb})), should be equal to $rV$ at $z=0$.
Then, the corresponding solution of the SAW is obtained by $\bm{U}_\mathrm{op}=\bm{U}' / U'_{4}(z=0) \times rV$, where $U'_{4}(z=0)$ denotes the value of the forth component of $\bm{U}'$ at $z=0$.
Since the first component of $\bm{U}_\mathrm{op}$, denoted as $U_{\mathrm{op},1}$, represents the lattice displacement along $x$ ($u_x$), the longitudinal strain $S_{xx}$ is derived by multiplying $iq$ to $U_{\mathrm{op},1}$. Figures~\ref{fig:Fig9_rev}(g) and \ref{fig:Fig9_rev}(h) shows $|S_{xx}|$ calculated using the values of $r$ in Figs.~\ref{fig:Fig9_rev}(e) and \ref{fig:Fig9_rev}(f) for $V=0.32$ V, which corresponds to an input rf power ($P$) of 0 dBm. Note that $V$ and $P$ are related by $P=|V|^2/2R_s$. $|S_{xx}|$ is in the range of $10^{-6}$ to $10^{-5}$, in agreement with previous studies~\cite{weiler2011prl,tateno2020prapp}.

Finally, we determine the SAW amplitude at the shorted boundary. 
Let $\bm{U}''$ be one of the plane-wave solutions of the SAW under the shorted bounday condition.
In this case, we cannot assume the same approach described above for the open boundary condition since the scalar potential $\phi$ vanishes at the substrate/perfect conductor interface. Instead, 
we assume that $u_x$ at $z=0$ does not change when the SAW travels from the substrate/vacuum to substrate/perfect conductor. This means that the SAW is not reflected at the film edge and no mechanical loss occurs in the film. If we adopt this assumption, the corresponding plane-wave solution is derived by $\bm{U}_\mathrm{sh} = \bm{U}'' / U''_{1} (z=0) \times  U_{\mathrm{op},1}(z=0)$, where $U''_{1}(z=0)$ and $U_{\mathrm{op},1}(z=0)$ denote the value of the first component of $\bm{U}''$ and $\bm{U}_\mathrm{op}$ at $z=0$, respectively. 

\section{Acknowledgments}
This work was supported by JSPS KAKENHI (Grant Number 20J20952, 20J21915, 23KJ1159, 23KJ1419, and 23H05463) from JSPS and MEXT Initiative to Establish Next-generation Novel Integrated Circuits Centers (X-NICS). H. M. acknowledges the JSR Fellowship from the University of Tokyo.

\bibliography{reference_JAP}

\clearpage

\end{document}